\documentclass[onecolumn,showpacs,preprintnumbers,amsmath,amssymb]{revtex4}
\usepackage{graphicx}% Include figure files
\usepackage{dcolumn}% Align table columns on decimal point
\usepackage{bm}% bold math
\usepackage{mathrsfs}
\usepackage{subfigure}

\begin{document}

\preprint{APS/123-QED}

\title{Mean First Hitting Time of Searching for Path \\ Through Random Walks on Complex Networks}% Force line breaks with \\

\author{Shao-Ping Wang}
 \affiliation{School of Information Technology, Jinling Institute of Technology, China}%Lines break automatically or can be forced with \\
 \email{spwang@jit.edu.cn}
\author{Wen-Jiang Pei}
\affiliation{School of Information Science and Engineering, Southeast University, China}%
\date{\today}% It is always \today, today,
             %  but any date may be explicitly specified

\begin{abstract}
We study the problem of searching for a fixed path $\epsilon_0\epsilon_1\cdots\epsilon_l$ on a network through random walks. We analyze the first hitting time of tracking the path, and obtain exact expression of mean first hitting time $\langle T \rangle$. Surprisingly we find that $\langle T \rangle$ is divided into two distinct parts: $T_1$ and $T_2$. The first part $T_1 =2m\prod_{i=1}^{l-1}d(\epsilon_i)$, is related with the path itself and is proportional to the degree product. The second part $T_2$ is related with the network structure. Based on the analytic results, we propose a natural measure for each path, i.e. $\varphi=\prod_{i=1}^{l-1}d(\epsilon_i)$, and call it random walk path measure(RWPM). $\varphi$ essentially determines a path's performance in searching and transporting processes. By minimizing $\varphi$, we also find RW optimal routing which is a combination of random walk and shortest path routing. RW optimal routing can effectively balance traffic load on nodes and edges across the whole network, and is superior to shortest path routing on any type of complex networks. Numerical simulations confirm our analysis.
\end{abstract}

% PACS, the Physics and Astronomy
                             % Classification Scheme.
%Use showkeys class option if keyword
                              %display desired
\maketitle

%%%%%%%%%%%%%%%%%%%%%%%%%%%%%%%%%%%%%%%%%%%%%%%%%%%%%%%%%%%%%%%%%%%%%%%%%%%%%%%%%%%%%

\section{Introduction}

Random walk is a fundamental dynamic process on graphs or networks. It has long been research interest in graph theory and related fields. Analysis on random walks has obtained fruitful results about hitting time,cover time, mixing rate and so on\cite{Bollobas1998}\cite{Lovasz1993}. Recent studies show that random walks on complex networks can reveal a variety of important features of the underlying network, such as diameter\cite{Lee2008}, centrality\cite{Newman2005}\cite{Guimera2002}, entropy\cite{Rosvall2005}, community structure\cite{Zhou2003}, topological structure\cite{Yoon2007}, etc. Random walk has also been exploited to tackle diffusing\cite{Lee2006}, searching\cite{Bisnik2007}, routing\cite{Tian2006}, immunization\cite{Hu2006}, traffic\cite{Germano2006} and communication\cite{Noh2004} problems on complex networks. All these research works show that random walk theory is a very useful tool  in the study of complex networks.  As is pointed by Newman\cite{Newman2003}, the ultimate goal of the study of complex networks is to discover governing laws of the workings of systems built upon those networks. Therefore, the real power of random walk theory lies in  disclosing basic properties of interaction between dynamical processes on complex networks and topological structure of these networks. In this letter, we go a step forward along this direction.

Apart from searching for a single node on a network, a random walker can also search for more complex targets. Here we study the problem of searching for a fixed path through random walks on complex networks\cite{spwang2009}. This is a phenomenon that can often be observed around us.
For example, many of us have the experience of running into an unexpected path on our journey.
The problem can be described as follows: Let $G(V,E)$ be a network  with node set $V$ and edge set $E$.
 There is a path $R$ on the network,
$R=\epsilon_0\epsilon_1\cdots\epsilon_l \;({\scriptstyle\epsilon_i\in V, \;0\leq i\leq l})$. $R$ is also denoted as $R(\epsilon_0\rightarrow \epsilon_l)$.
A random walker travels on the network, its $t$-step trace is
$W(t)=\sigma_0\sigma_1\cdots\sigma_t$ $({\scriptstyle\sigma_i\in V,\;0\leq i\leq t})$.
$W(t)$ is also called a $t$-step walk.
If the walker passes through path $R$ from its
initial node $\epsilon_0$ to its terminal node $\epsilon_l$ in successive steps, we say that the walker
finds path $R$, such an event can be expressed as $R\subseteq W(t)$. Let $T$ be the time that the walker finds path $R$
for the first time, i.e., $T=\textrm{min}\{\tau:R\subseteq W(\tau),\,\tau>0\}$.
By conventions here $T$ is still called first hitting time of random walks,
though it is now in the case of searching for a given path.
We will analyze first hitting time $T$,
and give formulas of the probability distribution of $T$, i.e. $f(t)=\textrm{Prob}\{X=t\}$,
 and mean first hitting time $\langle T \rangle$, i.e. $\langle T \rangle=\sum_{t=0}^{\infty}tf(t)$.
Based on the analytic results, we will propose a random walk path measure(RWPM) for each path and show that RWPM plays an important role in searching and transporting processes.

The outline of this paper is as follows. In Section II, we derive in detail the exact expressions of first hitting probability $f(t)$ and mean first hitting time $\langle T \rangle$. Surprisingly we find that $\langle T \rangle$ is divided into two distinct parts: $T_1$ and $T_2$. The first part $T_1$ is related with the path itself and the second part $T_2$ is related with the network structure. In Section III, we define random walk path measure(RWPM) based on our analytic results, and discuss its meaning and applications in searching and transporting processes. In Section IV, we give our conclusions.

\section{Analysis of First Hitting Time}

Now let's first proceed with the first hitting time problem.
 To simplify our discussion, we consider only
simple connected networks, where simple means without loops and multiple edges.
Let $|V|=n$ be the number of nodes, and $|E|=m$ be the number of edges.
 The adjacent matrix of the network is $\mathbf{A}=(a_{ij})_{n\times n}$, where $a_{ij}=1$ if  node $i$ and node $j$ are linked by an edge,  otherwise $a_{ij}=0$.
$d(i)$ is the degree of node $i$ , i.e. $d(i)=\sum_{j=1}^{n}a_{ij}$.
Suppose a random walker starts at source node $s$, and wanders on the network.
Each time it jumps with equal probability onto one of its neighboring nodes.
The transition probability matrix of the walker can be expressed as
$\mathbf{P}=\mathbf{D}\mathbf{A}$, where $\mathbf{D}$ is a diagonal
matrix, $\mathbf{D}=\textrm{diag}(1/d(1),...,1/d(n))$, and
$\mathbf{P}=(p_{ij})_{n\times n},\; p_{ij}=a_{ij}/d(i)$. The
probability that the walker goes from node $i$ to node $j$ in $m$
steps is $p_{ij}^{(m)}=\{\mathbf{P}^{m}\}_{ij}$.

A path $R$ on the network is a sequence of distinct nodes defined as above,
where $\epsilon_i\epsilon_{i+1}\,({\scriptstyle 0\leq i\leq l-1})$ are edges in $E$.
Let $u=\epsilon_0,\, v=\epsilon_l$, then $R=R(u \rightarrow v)$.
The length of $R$ is $l$. Without loss of generality, we assume $l\geq 1$.
If the walker finds path $R$, it tracks nodes of $R$ sequentially,
so there is a subwalk in $W(t)$ equal to $R$,
i.e., $R=\sigma_i\cdots\sigma_{i+l}\;({\scriptstyle 0\leq i,\;i+l\leq t})$,
which is denoted as $R\subseteq W(t)$. If the first hitting time $T=\tau$,
 then the walker tracks $R$ at step $\tau$ for the first time,
i.e., $R=\sigma_{\tau-l}\cdots\sigma_{\tau}$.
In the following, we will try to calculate the first hitting probability $f(t)=\textrm{Prob}\{T=t\}$.
However, it is not easy to compute $f(t)$ directly. We need some techniques.

Let's consider a conditioned probability $\theta(t)$. Without loss of generality, we assume $s\neq u$ in the proceeding discussion.  If the walker starting from source node $s$ arrives at $u$, the initial node of $R$, at step $t$, then what is
the probability of finding path $R$ in these $t$ steps? Such a probability can be expressed as
\begin{eqnarray}
\theta(t)&=&\textrm{Prob}\{T\leq t\mid\sigma_0=s,\;\sigma_t=u\}\nonumber\\
&=&\textrm{Prob}\{R\subseteq W_0(t)\}
\label{theta definition}
\end{eqnarray}
where $W_0(t)$ is a $t$-step walk starting from $s$ and arriving at $u$ at the last step.
 From eq.(\ref{theta definition}) it is easily seen that $\theta(t)=0\,\mbox{for}\;t=0,\ldots,l+1$.
Let $W_0^+(t)$ be a $W_0(t)$ that contains $R$, and $W_0^-(t)$ be a $W_0(t)$ that does not contain $R$.
 In order to calculate $\theta(t)$, we divide $W_0^+(t)$ into three subwalks:
\begin{equation}
\sigma_0\cdots\sigma_r,\quad
 \epsilon_0\cdots \epsilon_l,\quad
  \sigma_{r+l}\cdots \sigma_t
\label{ThreeSubwalks}
\end{equation}
where $\sigma_r=\epsilon_0, \epsilon_l=\sigma_{r+l}, \sigma_t=\epsilon_0$.
 The three subwalks in eq.(\ref{ThreeSubwalks}) also looks as
$s\cdots u,\, u\cdots v,\, v\cdots u $.
 The first subwalk starts from source node $s$ and arrives at $u$ at the last step, but it does not contain $R$. So the first subwalk can be written as $W_0^-(r)$.
Therefore, $\theta(t)$ can be computed by using iterations as below:
\begin{eqnarray}
\theta(t)&=&\textrm{Prob}\{X\leq t\mid \sigma_0=s,\sigma_t=u\}=\textrm{Prob}\{R\subseteq W_0(t)\}\nonumber\\
&=&\sum_{W_{0}^{+}(t)}p_{\sigma_0\sigma_1}\cdots p_{\sigma_{t-1}\sigma_t}\nonumber\\
&=&\sum_{r=1}^{t-l-1}\biggl[\sum_{W_{0}^{-}(r)}(
p_{\sigma_0\sigma_1}\cdots p_{\sigma_{r-1}\sigma_r})\cdot p_{\epsilon_0\epsilon_1}\cdots p_{\epsilon_{l-1}\epsilon_l}
 \cdot\sum_{\sigma_{r+l+1},\ldots,\sigma_{t-1}}
  (p_{\sigma_{r+l}\sigma_{r+l+1}}\cdots p_{\sigma_{t-1}\sigma_t})\biggr]\nonumber\\
&=&P_0\sum_{r=1}^{t-l-1}\biggl[\sum_{W_{0}^{-}(r)}(
p_{\sigma_0\sigma_1}\cdots p_{\sigma_{r-1}\sigma_r})
 \cdot\sum_{\sigma_{r+l+1},\ldots,\sigma_{t-1}}
  (p_{\sigma_{r+l}\sigma_{r+l+1}}\cdots p_{\sigma_{t-1}\sigma_t})\biggr]\nonumber\\
&=&P_0\sum_{r=1}^{t-l-1}\big(p_{su}^{(r)}-\theta(r)\big)\cdot p_{vu}^{(t-l-r)}
\label{theta}
\end{eqnarray}
with $\sigma_0=s,\sigma_r=u,\sigma_{r+l}=v,\sigma_{t}=u$, $P_0=p_{\epsilon_0\epsilon_1}\cdots p_{\epsilon_{l-1}\epsilon_l}$, and
\begin{equation}
p_{vu}^{(t-l-r)}=\sum_{\sigma_{r+l+1},\ldots,\sigma_{t-1}}p_{\sigma_{r+l}\sigma_{r+l+1}}\cdots
p_{\sigma_{t-1}\sigma_t}
\end{equation}
\begin{eqnarray}
&&\sum_{W_{0}^{-}(r)}
p_{\sigma_0\sigma_1}\cdots
p_{\sigma_{r-1}\sigma_r}\nonumber\\
&=&\sum_{W_{0}(r)}p_{\sigma_0\sigma_1}\cdots
 p_{\sigma_{r-1}\sigma_r}-\sum_{W_{0}^{+}(r)}
p_{\sigma_0\sigma_1}\cdots p_{\sigma_{r-1}\sigma_r}\nonumber\\
&=&p_{su}^{(r)}-\theta(r)
\end{eqnarray}
Let $\mathscr{P}_{ij}(x)$ be the generating function of
$\{p_{ij}^{(t)}\}_{t=0}^{\infty}$, and $\Theta(x)$ be the generating
function of $\{\theta(t)\}_{t=0}^{\infty}$, then $
\mathscr{P}_{ij}(x)=\sum_{t=0}^{\infty}p_{ij}^{(t)}x^t$ and $
\Theta(x)=\sum_{t=0}^{\infty}\theta(t)x^t$. Since $0\leq
p_{ij}^{(t)},\theta(t)\leq 1$, $\mathscr{P}_{ij}(x)$ and $\Theta(x)$
are convergent $\forall x\in(-1,1)$. By using the expression of t-step transition probability\cite{Lovasz1993}:
$p_{ij}^{(t)}=2m/d(j)+\sum_{k=2}^{n}\lambda_k^t\xi_{ki}\xi_{kj}\sqrt{d(j)/d(i)}$,\,
  we have
\begin{equation}
\mathscr{P}_{ij}(x)=\frac{2m}{d(j)(1-x)}+\sum_{k=2}^{n}\frac{1}{1-\lambda_kx}\xi_{ki}\xi_{kj}\sqrt{\frac{d(j)}{d(i)}}
\label{generating function of Pij}
\end{equation}
where $\lambda_k$ are eigenvalues of matrix
$\mathbf{Q}=\mathbf{D}^{-1/2}\mathbf{P}\mathbf{D}^{1/2}$, and
$\xi_k$ are corresponding eigenvectors.
Consequently, $\Theta(x)$ can be expressed as
\begin{eqnarray}
\Theta(x)&=&\sum_{t=0}^{\infty}\theta(t)x^t=\sum_{t=l+2}^{\infty}\theta(t)x^t\nonumber\\
&=&\sum_{t=l+2}^{\infty}P_0\sum_{r=1}^{t-l-1}\big(p_{su}^{(r)}-\theta(r)\big)\cdot p_{vu}^{(t-l-r)}\cdot x^t\nonumber\\
&=&P_0x^l\sum_{t=l+2}^{\infty}\sum_{r=1}^{t-l-1}\bigl(p_{su}^{(r)}-\theta(r)\bigr)\cdot x^r\cdot p_{vu}^{(t-l-r)}\cdot x^{t-l-r}\nonumber\\
&=&P_0x^l\sum_{k=1}^{\infty}\sum_{r=1}^{k}\big(p_{su}^{(r)}-\theta(r)\big)\cdot x^r\cdot p_{vu}^{(k-r+1)}\cdot x^{k-r+1}\nonumber\\
&=&P_0 x^l\sum_{i=1}^{\infty}\big(p_{su}^{(i)}-\theta(i)\big)\cdot x^i\cdot\sum_{j=1}^{\infty}p_{vu}^{(j)}\cdot x^j\nonumber\\
&=&P_0x^l\big[\mathscr{P}_{su}(x)-\Theta(x)\big]\cdot\mathscr{P}_{vu}(x)
\label{thetagfun}
\end{eqnarray}
therefore we have
\begin{equation}
\Theta(x)=\frac{P_0x^l\mathscr{P}_{su}(x)\mathscr{P}_{vu}(x)}
{1+P_0x^l\mathscr{P}_{vu}(x)}
\label{generating funtion of theta}
\end{equation}
Now we are ready to compute $f(t)$. Notice that $T=t$ means that the walker finds path $R$ for
the first time at exactly the $t$'th step. We denote such a walk as $W_1^+(t)$.  $W_1^+(t)$ can be divided into two subwalks: $\sigma_0\cdots\sigma_{t-l}\,,\,\epsilon_0\cdots\epsilon_l$,
with $\sigma_{t-l}=\epsilon_0$. $W_1^+(t)$ also looks as $s\ldots u,\,u\ldots
v$. The first subwalk does not contain $R$, so it can be denoted as $W_0^-(t-l)$. Note that $f(t)=0({\scriptstyle 0\leq t\leq l})$ since the walker can not
find path $R$ in less than $l$ steps. For $t>l$, $f(t)$ can be
computed as below:
\begin{eqnarray}
f(t)&=&\textrm{Prob}\{T=t\mid \sigma_0=s\}\nonumber\\
&=&\sum_{W_1^+(t)}(p_{\sigma_0\sigma_1}\cdots
p_{\sigma_{t-l-1}\sigma_{t-l}})\cdot (p_{\epsilon_0\epsilon_1} \cdots p_{\epsilon_{l-1}\epsilon_l})\nonumber\\
&=&\sum_{W_0^-(t-l)}(p_{\sigma_0\sigma_1}\cdots
p_{\sigma_{t-l-1}\sigma_{t-l}})\cdot P_0\nonumber\\
&=&\big[p_{su}^{(t-l)}-\theta(t-l)\big]\cdot P_0, \qquad (t>l)
\label{first hitting time}
\end{eqnarray}
By using eq.(\ref{generating funtion of theta}) and eq.(\ref{first hitting time}) the generating function of $\{f(t)\}_{t=0}^{\infty}$ can be derived
as below:
\begin{eqnarray}
\mathscr{F}(x)&=&\sum_{t=0}^{\infty}f(t)\cdot x^t=\sum_{t=l+1}^{\infty}f(t)\cdot x^t\nonumber\\
&=&\sum_{t=l+1}^{\infty}\big[p_{su}^{(t-l)}-\theta(t-l)\big]\cdot
P_0\cdot x^t\nonumber\\
&=&P_0x^l\sum_{t=l+1}^{\infty}\big[p_{su}^{(t-l)}-\theta(t-l)\big]\cdot
x^{t-l}\nonumber\\
&=&P_0x^l\big[\mathscr{P}_{su}(x)-\Theta(x)\big]\nonumber\\
&=&\frac{P_0x^l\mathscr{P}_{su}(x)}{1+P_0x^l\mathscr{P}_{vu}(x)}
\label{generating function of f(t)}
\end{eqnarray}
$f(t)$ can be calculated by taking derivatives of $\mathscr{F}(x)$, i.e.,
$f(t)=\frac{1}{t!}\,\frac{d^{t}}{dx^{t}}\;\mathscr{F}(x)\,|_{x=0}$.
At the same time, by using eq.(\ref{generating function of Pij}) and eq.(\ref{generating function of f(t)}), mean first passage time $\langle T\rangle$ can be calculated as
the following:
\begin{eqnarray}
&&\langle T\rangle=\sum_{t=0}^{\infty}t f(t)=\mathscr{F}'(1)=\lim_{x\rightarrow 1}\mathscr{F}'(x)\nonumber\\
&=&\lim_{x\rightarrow 1}\frac{1}{\big[1+P_0x^l\mathscr{P}_{vu}(x)\big]^2}
\Big\{P_0\big[lx^{l-1}\mathscr{P}_{su}(x)+x^l\mathscr{P}'_{su}(x)\big]\cdot\big[1+P_0x^l\mathscr{P}_{vu}(x)\big]\nonumber\\
&&-P_0x^l\mathscr{P}_{su}(x)\cdot P_0\big[lx^{l-1}\mathscr{P}_{vu}(x)+x^l\mathscr{P}'_{vu}(x)\big]\Big\}\nonumber\\
&=&\frac{2m}{d(u)P_0}+2m\sum_{k=2}^{n}\frac{1}{1-\lambda_k}\Big[\frac{\xi_{kv}\xi_{ku}}{\sqrt{d(v)d(u)}}-\frac{\xi_{ks}\xi_{ku}}{\sqrt{d(s)d(u)}}\Big]\nonumber\\
&=&2m\prod_{i=1}^{l-1}d(\epsilon_i)+
2m\sum_{k=2}^{n}\frac{1}{1-\lambda_k}\Big[\frac{\xi_{kv}\xi_{ku}}{\sqrt{d(v)d(u)}}-\frac{\xi_{ks}\xi_{ku}}{\sqrt{d(s)d(u)}}\Big]
\label{mean hitting time}
\end{eqnarray}
Up to now, we have obtained main results of this paper, i.e., the probability distribution $f(t)$ and mean first hitting time $\langle T\rangle$.  From eq.(\ref{mean hitting time}) we see that the mean first hitting time is divided into two parts, $\langle T\rangle=T_1+T_2$. The first part $T_1$ is $2m\prod_{i=1}^{l-1}d(\epsilon_i)$, which is proportional to the product of degrees of nodes on path $R$ except the initial and terminal nodes. The second part $T_2$ is more complicated since it includes information of topological structure of the network. But if we fix source node $s$, initial node $u$ and terminal node $v$, then the second part $T_2$ is constant for all paths between $u$ and $v$, and also independent of path length $l$. This is a somewhat surprising result. Nevertheless it gives us a standpoint to evaluate and compare all paths between $u$ and $v$. In fact, the difference only comes from the first part, or simply $\prod_{i=1}^{l-1}d(\epsilon_i)$, the product of degrees of nodes on the path. Judging by our intuitions, we know that this is a reasonable result.

\section{Random Walk Path Measure}

Based on the above analysis, we define a function $\varphi(R(\epsilon_0\rightarrow \epsilon_l))=\prod_{i=1}^{l-1}d(\epsilon_i)$ for path $R$. In comparing all paths between two points $u$ and $v$, $\varphi$ has the same effect as the mean first hitting time $\langle T\rangle$. That is to say, a path with small $\langle T\rangle$ will also has small $\varphi$, vice versa. Therefore, $\varphi$ is a natural measure associated with each path. We call it random walk path measure(RWPM). In fact, RWPM determines a path's performance in searching and transporting processes. In the following, we use two examples to make this point clear.

\begin{figure}[h]
\centering
\subfigure[]
{
    \label{fig1:a} %% label for first subfigure
    \includegraphics[width=65mm]{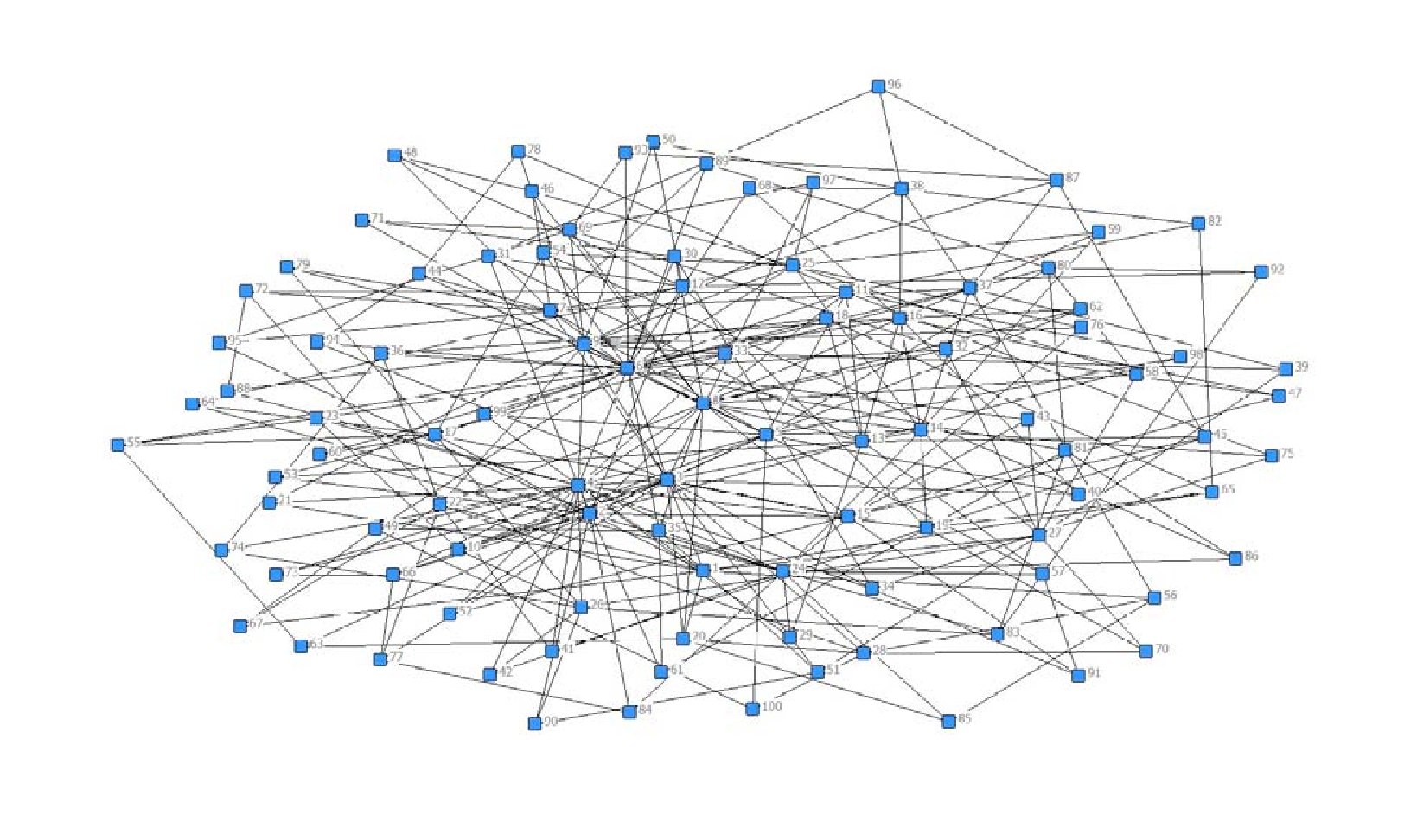}
}
\hfill
\subfigure[]
{
    \label{fig1:b} %% label for second subfigure
    \includegraphics[width=65mm]{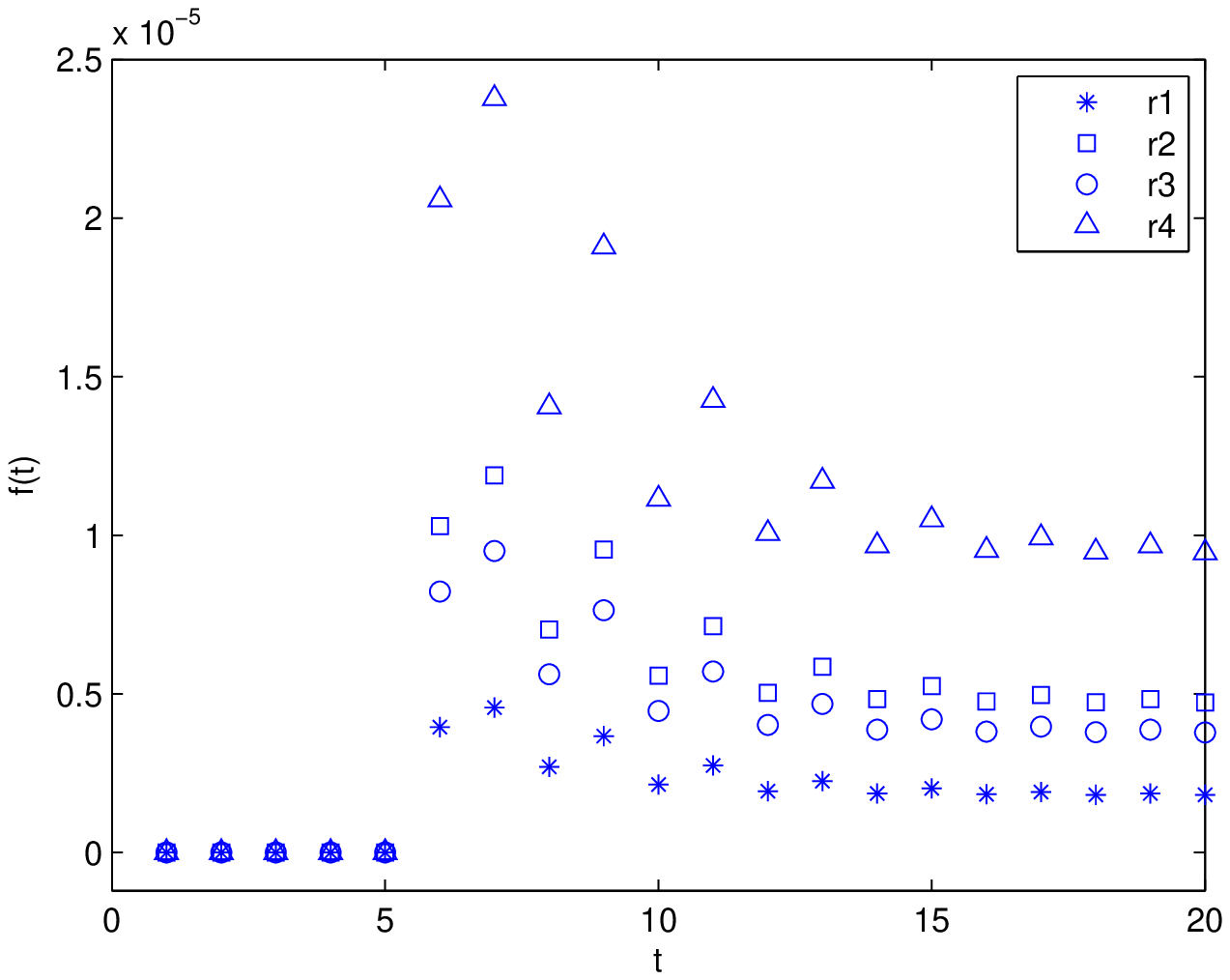}
}
\caption {
Comparing shortest paths by their first hitting probability.
(a) A simple scale-free network with 100 nodes generated by using BA model\cite{Barabasi99}.
(b) Distribution of first hitting probability of four shortest paths between a pair of nodes.
The information of the nodes on these paths is list in TABLE \ref{tab:table1}.
}
\label{fig1:compare shortest paths} %% label for entire figure
\end{figure}

The first example shows that paths between a pair of nodes have different performance in searching processes according to their RWPMs. There are four shortest paths between node 70 and node 96 in a small network(Fig.\ref{fig1:a}), they have the same initial node $u=70$ and the same terminal node $v=96$. Details of these paths are listed in TABLE \ref{tab:table1}. $\varphi_4=180$ is the smallest and $\varphi_1=936$ is the largest. In Fig.\ref{fig1:b} we see that distribution of their first hitting probability are different, route4 has the largest first hitting probability all the time while route1 has the smallest first hitting probability.  That is to say, it is easy to search for route4 and difficult to search for route1 if a random walker starts from same source node $s$. Here we see that searching efficiency of all paths between $u$ and $v$ are determined by their RWPM, the smaller the RWPM of a path, the more easily it will be found. By the way, we also point out here that shortest paths can be further distinguished by comparing their RWPMs.

 %compare first hitting probability of shortest paths

\begin{table}
\caption{
Four shortest paths on a small BA network(Fig.\ref{fig1:a}). For example, router1 is 70-19-4-31-96. The initial node is 70 and the terminal node is 96. The path length is 4. Node degree is shown in brackets. The last column is $\varphi$ value.
}
\begin{ruledtabular}
\begin{tabular}{lllllll}
          & v0    & v1    & v2     & v3    & v4    & $\varphi$\\
\hline
   route1 & 70(3) & 19(9) & 4(26)  & 31(4) & 96(3) & 936\\
   route2 & 70(3) & 19(9) & 12(10) & 31(4) & 96(3) & 360\\
   route3 & 70(3) & 19(9) & 12(10) & 87(5) & 96(3) & 450\\
   route4 & 70(3) & 19(9) & 45(4)  & 87(5) & 96(3) & 180\\
\end{tabular}
\end{ruledtabular}
\label{tab:table1}
\end{table}
%%

% Why is so? It is not difficult to answer this question in view of their searching information\cite{Rosvall2005}.  From information theory we know that information is a measure of uncertainty. Then searching information represents uncertainty for a random walker remaining on the right path. In a network, the uncertainty comes from branches at each node. The more branches a node has, the more uncertainty it contributes to the path it lies on, then more possible it is for a walker to deviate from the right direction. On a given path, it is degree product $\prod_{i=0}^{l-1}d(\epsilon_i)$ instead of single node's degree that determines overall uncertainty. In searching for path $R(u\rightarrow v)$, a random walker needs $\log_2 [d(u)\varphi]$ bits of information. Correspondingly, paths with small mean first hitting time will need less searching information than those with large mean first hitting time. So they are easy to be found.

\begin{figure}[h]
\centering
\subfigure[]
{
    \label{fig2:a} %% label for first subfigure
    \includegraphics[width=65mm]{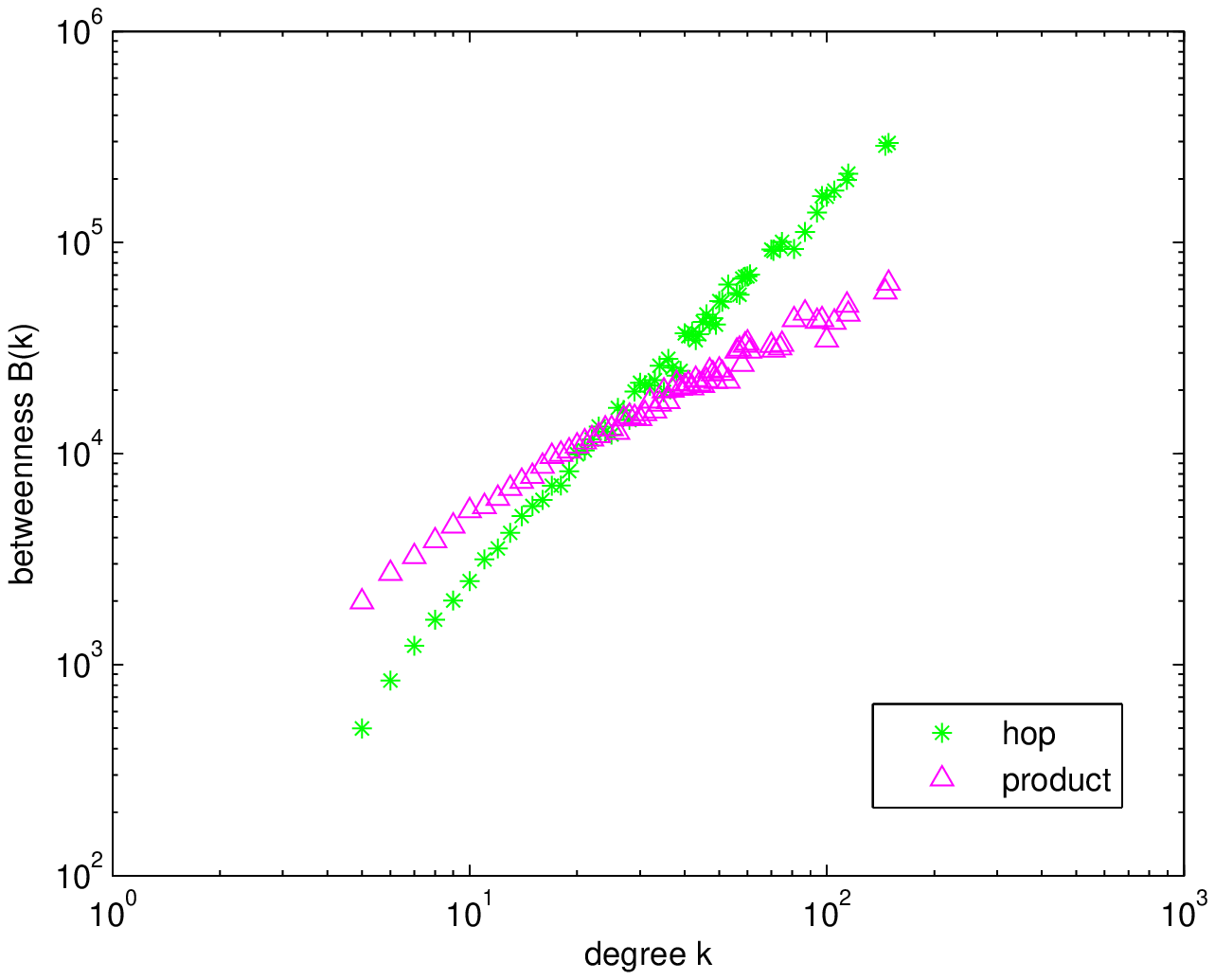}
}
\hfill
\subfigure[]
{
    \label{fig2:b} %% label for second subfigure
    \includegraphics[width=65mm]{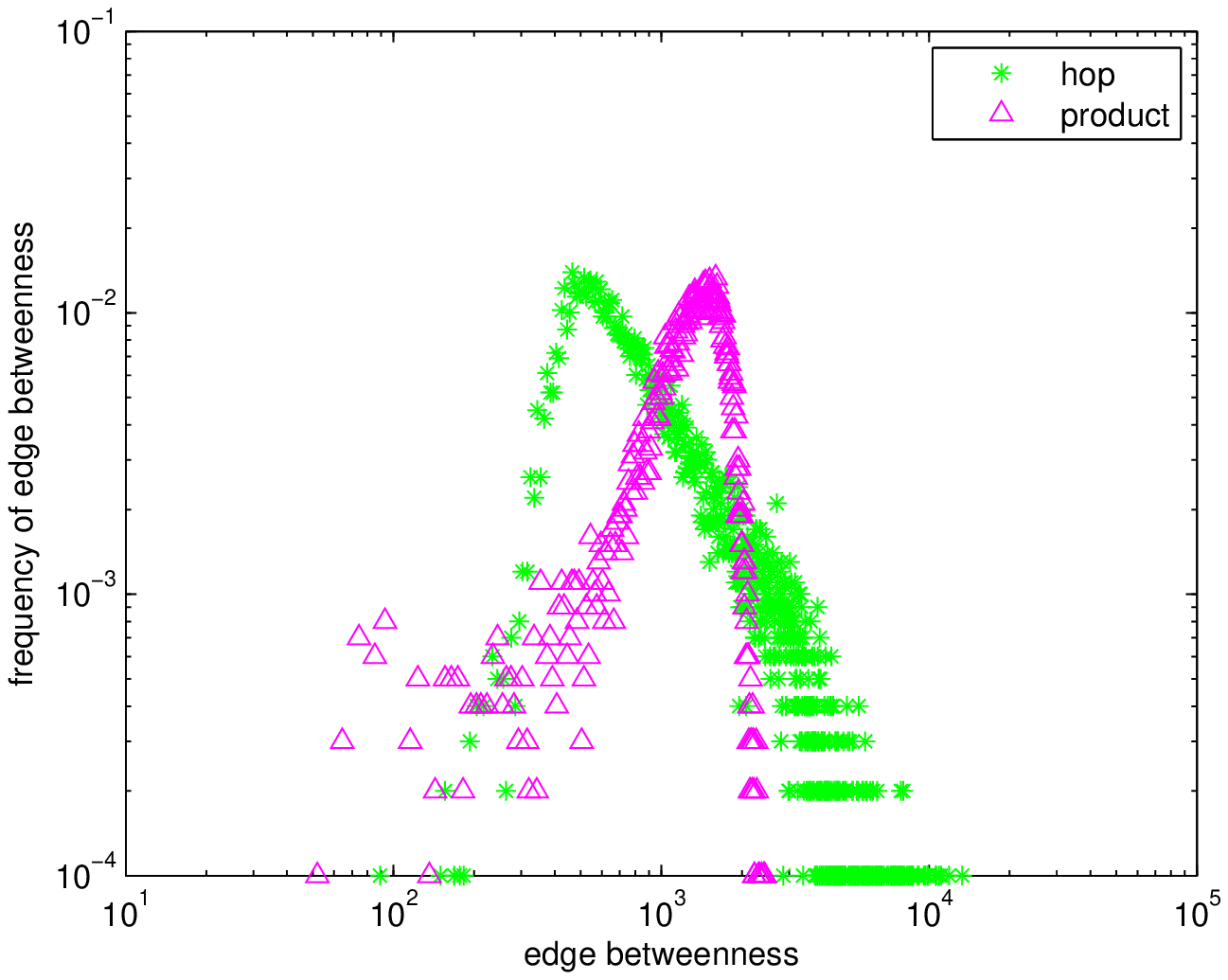}
}
\caption {
Node betweenness and edge betweenness for shortest path routing (hop) and RW optimal routing (product) on a BA scale-free network with 2000 nodes.
(a) Node betweenness $B(k)$ vs node degree $k$. Node betweenness is averaged according to node degree. (b) Frequency of edge betweenness.
}
\label{fig2:betweenness} %% label for entire figure
\end{figure}

 The second example shows that RWPM can be applied in transportation to design more efficient routing strategy. In recent years, the issue of transportation on complex networks attracted a lot of attention from researchers in this field\cite{Chen2006}\cite{Echenique2004}\cite{Danila2006}\cite{Scholza2008}\cite{Wang2006a}\cite{Wang2006b}\cite{Tadic2004}.
 One of the most important problems in transportation is to
 balance load on nodes and edges across the whole network so as to enhance the network's overall transporting capacity.
 This can be achieved by designing better routing strategies that are able to undo correlation between traffic load and structure of the network.
 Some efforts have been made towards this direction. Danila {\it et al} proposes an extreme optimization algorithm to  balance traffic on a network by minimizing the maximum  node betweenness\cite{Danila2006}.
 Scholz {\it et al} suggests smoothing techniques to distribute load evenly across the network\cite{Scholza2008}. However, their algorithms are not practical for designing routing strategy because of either overwhelming computational cost or unstable output.
 Theoretically random walks can accomplish perfect decorrelation between traffic load and network structure since in steady state, a random walker visits any node $i$ with probability $d(i)/2m$, proportional to the node's degree, and visits any edge  with equal probability $1/2m$\cite{Lovasz1993}. However, random walk can not produce stable routing.  Hence we propose a new routing strategy by using paths with smallest RWPM, and we call it RW optimal routing or RW optimal path. In other words, among all paths between a pair of nodes $u$ and $v$, RW optimal paths are those making $\varphi$ minimum.
  Such a routing strategy naturally inherits merits from both random walk and shortest path routing.
  On one hand, RW optimal path has minimum $\varphi$ value, so it can not be too long.
  In fact, RW optimal path is only slightly longer than shortest path on average, for example, in a BA network illustrated in Fig.\ref{fig2:betweenness}, the average path length is 3.2 for shortest paths while the average path length is 3.4 for RW optimal paths.
  On the other hand, minimum degree product also ensures that RW optimal paths are the least possible to be interfered from outside since they have minimum branches linking other parts of the network. Therefore, among all paths between a pair of nodes, RW optimal path is the least possible to get into congestion.

  Fig.\ref{fig2:betweenness} shows simulation results on a BA scale-free network. Fig.\ref{fig2:a} is distribution of node betweenness and Fig.\ref{fig2:b} is distribution of edge betweenness. The betweenness centrality is calculated by the method introduced in \cite{Yan06}.
  In Fig.\ref{fig2:a} we see that node betweenness has the form $B(k)\thicksim k^{\beta}$\cite{Goh2001}, $\beta\thickapprox 0.95$ for RW optimal routing and $\beta\thickapprox 1.82$ for shortest path routing. The node betweenness for RW optimal routing is close to a linear function of degree $k$. We also note that the largest node betweenness is 64138 for RW optimal routing while the largest node betweenness is 295948 for shortest path routing, The former is only one fifth of the latter. This is the first sign that RW optimal routing undo correlation between traffic load and network structure\cite{Scholza2008}. In Fig.\ref{fig2:b} edge betweenness are distributed much more narrowly for RW optimal routing than for shortest path routing. In the same time, largest edge betweenness is much smaller for RW optimal routing than for shortest path routing. In fact, the largest edge betweenness is 2431 for RW optimal routing while the largest edge betweenness is 13303 for shortest path routing. This is the second sign that RW optimal routing undo correlation between traffic load and network structure\cite{Scholza2008}. All these manifest that traffic load is distributed more evenly by using RW optimal routing than by using shortest path routing.
  We also make simulations on other type of networks, such as random graph, scale-free networks with tunable clustering\cite{Holme2002}, internet alike networks\cite{ShiZhou2004}, and so on,  and get similar results. In all the cases, RW optimal routing is superior to shortest path routing.

\section{Conclusions}

In summary, we have studied the problem of searching for path on complex networks through random walks and derived formulas for
 the distribution of first hitting time $f(t)$ and mean first hitting time $\langle T \rangle$. We find that $\langle T \rangle$ is  divided into two distinct parts with totally different features: the first part $T_1 $ is proportional to the product of the degrees of nodes on the path, i.e., $T_1 =2m\prod_{i=1}^{l-1}d(\epsilon_i)$, and the second part $T_2$ is determined by the topological structure of the network. Based on the analytic results, we define a function $\varphi$ for each path $R$, $\varphi(R(\epsilon_0\rightarrow \epsilon_l))=\prod_{i=1}^{l-1}d(\epsilon_i)$. $\varphi$ is called random walk path measure(RWPM). Since it is derived from the mean first hitting time, $\varphi$ is a natural measure associated with each path. We have shown that $\varphi$ essentially determines a path's performance in searching and transporting processes. By minimizing $\varphi$ we can also find RW optimal routing. RW optimal routing inherits merits from both random walk and shortest path routing. So it is superior to shortest path routing in all cases. Numerical simulations confirm our conclusion.

%% References
%%
%% Following citation commands can be used in the body text:
%% Usage of \cite is as follows:
%%   \cite{key}         ==>>  [#]
%%   \cite[chap. 2]{key} ==>> [#, chap. 2]
%%

%% References with bibTeX database:

\bibliographystyle{elsarticle-num}
%%\bibliography{<your-bib-database>}
\bibliography{MFHTofSearchPath}

%% Authors are advised to submit their bibtex database files. They are
%% requested to list a bibtex style file in the manuscript if they do
%% not want to use elsarticle-num.bst.

%% References without bibTeX database:

% \begin{thebibliography}{00}

%% \bibitem must have the following form:
%%   \bibitem{key}...
%%

% \bibitem{}

% \end{thebibliography}

\end{document}